\documentclass[11pt]{article}

\usepackage{amsmath}
\usepackage{amssymb}
\usepackage{graphicx}
\usepackage{subfigure}

\def\middlespace {\smallskipamount=5.625pt plus1.5pt minus1.5pt
                  \medskipamount=11.25pt plus3pt minus3pt
                  \bigskipamount=22.5pt plus6pt minus6pt
                  \normalbaselineskip=22.5pt plus0pt minus0pt
                  \normallineskip=1pt
                  \normallineskiplimit=0pt
                  \jot=5.625pt
                  {\def\smallskip {\vskip\smallskipamount}}
                  {\def\medskip   {\vskip\medskipamount}}
                  {\def\bigskip   {\vskip\bigskipamount}}
                  {\setbox\strutbox=\hbox{\vrule
                    height15.75pt depth6.75pt width 0pt}}
                  \parskip 11.25pt
                  \normalbaselines}

\begin{document}

\ \vskip 1.0 in

\begin{center}
 { \Large {\bf The connection between `emergence of time from quantum gravity' and}}

\smallskip

{\Large {\bf  `dynamical collapse of the wave-function in quantum mechanics'}}

\vskip 0.2 in

\smallskip

\bigskip

\bigskip

\bigskip

{{\large
{\bf Tejinder P. Singh 
} 
}}

\medskip

{\it Tata Institute of Fundamental Research,}\\
{\it Homi Bhabha Road, Mumbai 400 005, India}\\
\smallskip
{\tt e-mail address: tpsingh@tifr.res.in}\\
\bigskip
{\it 22nd March, 2010}
\vskip 0.5cm
\end{center}

\vskip 1.0 in

\begin{abstract}

\noindent There are various reasons to believe that quantum theory could be an emergent phenomenon. Trace Dynamics is an underlying classical dynamics of non-commuting matrices, from which quantum theory and classical mechanics have been shown to emerge, in the thermodynamic approximation. However, the time that is used to describe evolution in quantum theory is an external classical time, and is in turn expected to be an emergent feature - a relic of an underlying theory of quantum gravity. In this essay we borrow ideas from Trace Dynamics to show that classical time is a thermodynamic approximation to an operator time in quantum gravitational physics. This prediction will be put to test by ongoing laboratory experiments attempting to construct superposed states of macroscopic objects.  

\noindent 

\vskip 1.0 in

\centerline{\it This essay received an Honorable Mention in the Gravity Research Foundation Essay Competition 2010}
\centerline{\it To appear in Int. J. Mod. Phys. D [Dec. 2010 Special Issue]}     
\end{abstract}

\newpage

\middlespace

\noindent 

\noindent 
When one attempts to explain the outcome of a quantum measurement as being due to a possible dynamical collapse of the wave-function of the quantum system,  one assumes an external classical time as given.  At a deeper level though, we must also understand how this classical time emerges from quantum gravity. In this essay we show that the physics behind the emergence of classical time from quantum gravity is the same as the physics behind dynamical collapse of the wave-function. We  also suggest how this idea can be tested in the laboratory. 

\bigskip

\centerline{\it Dynamical collapse of the wave-function}

There are various reasons why our present understanding of quantum mechanics could be considered incomplete. These include the {\it ad hoc} introduction of probabilities in order to explain quantum measurement, and the fact that one must first know the classical theory before one can obtain its more general quantum version. It is desirable to have an underlying description of quantum theory, which is inherently quantum and deterministic, and from which standard quantum mechanics, probabilities, and classical mechanics, emerge in suitable approximations. The so-called `Trace Dynamics' constructed by Adler and collaborators is a remarkable theory which goes a long way towards achieving these goals \cite{adler}. Below we give a very brief summary of this theory, and in the next section we show how this work can be generalized to explain the emergence of time from quantum gravity.

Trace dynamics is the classical dynamics of $N\times N$ matrices (equivalently, operators) $q_r$, whose elements can be either complex numbers [bosonic matrices] or Grassman numbers [fermionic matrices]. The Lagrangian ${\bf L} [\{q_r\},\{\dot{q_r}\}]$ is the trace of a polynomial function of the matrices $\{q_r\}$ and their time derivatives $\{ \dot{q_r} \}$. The derivative with respect to an operator ${\cal O}$, of the trace 
${\bf P}$ of a polynomial $P$ made out of non-commuting operators is  defined as follows
\begin{equation}
\delta TrP = Tr \frac {\delta Tr P}{\delta {\cal O}}\delta{\cal O}.
\end{equation}
By proceeding as in ordinary classical mechanics one constructs an action; Lagrangian dynamics is derived from an action principle, and conjugate momenta $p_r$, a trace Hamiltonian ${\bf H}$ and Hamilton's equations of motion are constructed. Apart from the trace Hamiltonian there are two other important conserved quantities. One is the `trace fermionic number' $N\equiv \sum_F q_r p_r$ obtained by summing over fermionic variables. The other is the remarkable traceless and anti-self-adjoint Adler-Millard constant
\begin{equation}
\tilde{C}\equiv \sum_B [q_r,p_r]  - \sum_F \{q_r,p_r\}
\label{adler-miller}
\end{equation}
which is a result of the invariance of the Lagrangian under global unitary transformations of the $q_r$ and $p_r$. It is profound that such a 
conserved commutator should appear in a classical theory in which the matrices take arbitrary values. This plays a key role in the subsequent emergence of quantum mechanics. 

Since the dynamical variables $q_r,p_r$ take arbitrary values in phase space, a probability distribution $\rho$ is constructed for them, by maximizing the entropy of the canonical ensemble, subject to the constancy of the conserved quantities. The result is
\begin{equation}
\rho=Z^{-1}\exp-Tr[\tilde{\lambda}\tilde{C}+\tau{\bf H} + \eta {\bf N}]
\end{equation}
where $\tau$, $\eta$ and the traceless anti-self-adjoint matrix $\tilde{\lambda}_{ij}$ are Lagrange multipliers. An equipartition theorem
\begin{equation}
0=\int d\mu \ \delta(\rho A)/\delta x_r
\end{equation}
holds, where $d\mu$ is the volume measure in phase space, $x_r$ is $q_r$ or $p_r$, and $A$ is any polynomial function of the dynamical variables.
The mean value $<A>\equiv \int d\mu \ \rho A$ is a function only of $\tilde{\lambda}$, and in the basis in which $\tilde{\lambda}$ is diagonal, the average $<\tilde{C}>$ of the Adler-Millard constant is also diagonal. All the diagonal components of $<\tilde{C}>$ are assumed to be equal and have a value $i\hbar$ or $-i\hbar$. Further, assuming that $\tau^{-1}$ is the Planck temperature, one shows using the equipartition theorem that at low energies (compared to Planck energy) the
canonical distribution is governed by $\tilde{C}$, and not by ${\bf H}$. The 
 canonical average of the time evolution of  $x_r$ obeys the standard Heisenberg equations of motion, and the standard commutation relations of quantum theory also follow. From the Heisenberg equations one constructs a relativistic quantum field theory, and also obtains a non-relativistic Schr\"{o}dinger equation in the limit. 
 
Trace dynamics also suggests that during a quantum measurement the wave-function of the quantum system does indeed collapse, and that this happens in accordance with the Born probability rule. In the theory described above, quantum mechanics results from a thermodynamic limit of the underlying statistical mechanics, when $\tilde{C}$ is replaced by its average after using the equipartition theorem. To the next level of accuracy, one must take into account the Brownian motion fluctuations of $\tilde{C}$, around its average. This has the effect that the Schr\"{o}dinger equation acquires a stochastic non-linear correction. Assuming the fluctuation $f(t)$ in $\tilde{C}$ to be self-adjoint (a generalisation which is possible, and necessary for dynamical collapse) the modified Schr\"{o}dinger equation takes the form
\begin{equation}
i\hbar \frac{\partial\Psi}{\partial t} = H \Psi  -  i{\cal M}(t) \Psi    
\end{equation}
where the anti-self-adjoint correction term $i\cal{M}$ is determined by the fluctuation $f(t)$. The correction term can be specialized to the form assumed in the phenomenological Continuous Spontaneous Localization (CSL) model \cite{csl}, where the fluctuation is assumed to be of the white noise type. This causes a superposition of states to evolve to one of the states; which state will survive depends on the particular white noise function. The probability of outcome of a particular state is the same as the net probability of the white noise functions which result in this state and is equal to the square of the magnitude of this state's coefficient in the initial superposition. Superluminal signalling is avoided because the non-linearity is stochastic. Furthermore, it can be shown that for microscopic systems the corrections are negligible. The theory also explains why macroscopic objects are localized in position and not found in superposition of position states - the superposition lifetime is too small to be observed.

\bigskip

\centerline{\it Emergence of time from quantum gravity}

In the above theory time is treated as an external classical variable. However, there must exist an equivalent reformulation of quantum theory which does not refer to an external classical time. This is because classical time is part of a spacetime manifold on which there resides a classical spacetime metric. The metric is determined by classical matter fields, and in the [entirely plausible, in principle] absence of the latter, the former will undergo quantum fluctuations, which, in accordance with the Einstein hole argument, destroy the underlying spacetime manifold. Hence the necessity for the reformulation \cite{singh1}. Below, we demonstrate, using the ideas of Trace Dynamics, how classical time emerges from the apparently timeless theory having only quantum matter fields and a fluctuating spacetime \cite{singh2}. 

Assume that in the reformulation, spacetime is described by non-commuting coordinates $(\hat{x},\hat{y},\hat{z},\hat{t})$ which are operators (equivalently, matrices). If the total mass-energy of the particles in this spacetime is much less than Planck energy, the non-commutative line-element is assumed to be of the `Minkowski' form
\begin{equation}
ds^{2} = Tr [ d\hat{t}^{2} - d\hat{x}^{2} + d\hat{t}d\hat{x} - d\hat{x} d\hat{t} ]
\end{equation}
with obvious generalisation to the 4-d case. [Inclusion of gravity raises no conceptual difficulties for the argument that follows]. A non-commutative special relativity is constructed by defining an operator four-velocity
$\hat{u}_r^{i}=d\hat{q}_r^{i}/ds$ and an operator four-momentum $\hat{p}_r^{i}=m_r \hat{u}_r^{i}$ for a particle with four-position $\hat{q}_r$ and having mass $m_r$, with the condition $Tr[\hat{p}^{i}\hat{p}_{i}]=m_r^{2}$ being satisfied naturally. Assuming the existence of an auxiliary time-parameter $\tau$, and assuming that a particle is either bosonic or fermionic, one can construct Trace Lagrangian dynamics as before,
using ${\bf L} [\{q_r\},\{d{q_r}/d\tau \}]$. The central difference is that now the `time-location' $\hat{t}_r$ of every particle is an operator, and phase space is augmented by the conjugate pair $(\hat{t}_r,\hat{E}_r)$ for every particle. 

As before, global unitary invariance of the Lagrangian implies the existence of the Adler-Millard constant (\ref{adler-miller}), with the understanding that commutators/anti-commutators of $\hat{t}_r$ and $\hat{E}_r$ are included in $\tilde{C}$, and constancy is in evolution with respect to $\tau$. A statistical mechanics is constructed, and in the thermodynamic limit, the equipartition theorem implies the existence of 
commutation/anti-commutation relations and a Schr\"{o}dinger-like equation
\begin{equation}
i\hbar \frac{\partial \Psi}{\partial \tau}  = H \Psi.
\label{slike}
\end{equation}
Again it is clear that the configuration variables include the canonical pairs $(\hat{t}_r,\hat{E}_r)$. Indeed it is apparent that this is not ordinary quantum mechanics, but a generalisation when there is no external physical time. 

We now come to the punchline of this essay. Brownian motion fluctuations around the thermodynamic canonical average must be taken into account. If the Universe is dominated by macroscopic objects (such as today's Universe is), then, as in the CSL model, the stochastic non-linear corrections cause localization, though now not only in space, but also in {\it time}! Quantum fluctuations in time get suppressed, the commutator/anti-commutator between $\hat{E}_r$ and $\hat{t}_r$ vanishes \textit{and the auxiliary time parameter $\tau$ can be identified with classical time}. Given this background spacetime, which is dominated and determined by classical matter fields, and which provides an external classical time via
\begin{equation}
\hat{x}\rightarrow x, \quad \hat{t}\rightarrow t = \tau, \quad (d\hat{t}d\hat{x}-d\hat{x}d\hat{t})\rightarrow 0
\end{equation}
one can construct the ordinary quantum mechanics of elementary particles, as well as Adler's Trace Dynamics. 

There is a serious ongoing world-wide experimental effort to test whether or not one can prepare superposition of two quantum states of a macroscopic object \cite{expts}. If the answer is found to be yes, then quantum mechanics is valid at all scales, and dynamical collapse of the wave-function does not take place. If the answer is found to be no, dynamical collapse is strongly indicated. This could be an experimental verification for the CSL model and for Trace Dynamics, and hence, indirectly, also for our proposal that classical time emerges when Brownian motion fluctuations destroy the quantum nature of spacetime. This would be yet another intriguing example of a connection between statistical mechanics, thermodynamics, quantum theory and gravity. 

It is a pleasure to thank Kinjalk Lochan for helpful discussions.

\end{document}